# Exact Diagonalization of Heisenberg $SU(N)$ models

Pierre Nataf, Frédéric Mila[1]

[1]*Institut de Physique Théorique, École Polytechnique Fédérale de Lausanne (EPFL), CH-1015 Lausanne, Switzerland*
(Dated: August 22, 2014)

Building on advanced results on permutations, we show that it is possible to construct, for each irreducible representation of $SU(N)$, an orthonormal basis labelled by the set of *standard Young tableaux* in which the matrix of the Heisenberg $SU(N)$ model (the quantum permutation of $N$-color objects) takes an explicit and extremely simple form. Since the relative dimension of the full Hilbert space to that of the singlet space on $n$ sites increases very fast with $N$, this formulation allows to extend exact diagonalizations of finite clusters to much larger values of $N$ than accessible so far. Using this method, we show that, on the square lattice, there is long-range color order for $SU(5)$, spontaneous dimerization for $SU(8)$, and evidence in favor of a quantum liquid for $SU(10)$.

There is currently a considerable experimental activity on ultra cold multi-component fermions[1–3]. When loaded in an optical lattice, these systems are expected to be, at integer filling and for sufficiently large on-site repulsion, in a Mott insulating phase described by the $SU(N)$ Heisenberg model[4–7]. This effective model is a generalization of the familiar $SU(2)$ model, and in the case of one particle per site, it takes the general form of a quantum permutation Hamiltonian:

$$H = \sum_{(i,j)} J_{ij} \sum_{\mu,\gamma=A,B,C...} |\mu_i \gamma_j\rangle\langle\gamma_i \mu_j| = \sum_{(i,j)} J_{ij} P_{ij}, \quad (1)$$

where the sum $\sum_{(i,j)}$ runs over all pairs of interacting sites ($J_{ij}$ being the coupling constant). The *permutation* operator $P_{ij}$ simply switches the states between sites $i$ and $j$, and the local Hilbert space is of dimension $N$.

In the context of condensed matter physics, two cases have been mainly studied: $SU(3)$, which describes spins 1 with a biquadratic interaction equal to the bilinear one[8–10], and $SU(4)$, the symmetric version of the Kugel-Khomskii spin-orbital model[11–14]. Apart from one dimension, where there is a Bethe ansatz solution[15] and minus sign free Quantum Monte Carlo simulations[16, 17], reliable information could only be obtained by combining approximate analytical and numerical approaches such as flavor-wave theory[8, 18], exact diagonalizations (ED) of finite clusters[9, 19–21], variational Monte-Carlo[20–22] or tensor network algorithms[19, 20, 23].

With cold atoms, one can simulate larger values of $N$, up to 10, allowing one to realize new types of quantum phases[7]. In particular, it has been predicted by mean-field theory that chiral phases might be stabilized for large enough $N$[24–26]. However, for large $N$, most of the methods encounter specific difficulties: flavor-wave theory is limited to phases with long-range color order, the performance of tensor-network algorithms significantly decreases when the dimension of the local Hilbert space increases, and ED are severely limited by the size of the available clusters. Alternatives are clearly called for.

In this letter, we introduce a simple method to perform ED of any quantum permutation Hamiltonian separately in each *irreducible representation* (irrep) of $SU(N)$. Since the dimension of the irreps relevant at low energy (for instance the singlet, to which the ground state belongs) is much smaller than that of the sector used in traditional ED, this approach allows one to perform ED on essentially the same cluster sizes for large $N$ as for small $N$. The power of the method is illustrated by the first ED investigation so far of $SU(5)$, $SU(8)$ and $SU(10)$ on the square lattice.

Let us first recall some standard results about the irreps of $SU(N)$. For a lattice of $n$ sites, each irrep can be associated to a *Young tableau* with $n$ boxes and at most $N$ rows (see Fig.1). The shape of a Young tableau can be described by an array $\alpha = [\alpha_1, \alpha_2, ..., \alpha_k]$ ($1 \leq k \leq N$) where the lengths of the rows $\alpha_j$ satisfy $\alpha_1 \geq \alpha_2 \geq ... \geq \alpha_k \geq 1$. In the full Hilbert space $\square^{\otimes n}$, where $\square$ is the fundamental irrep or equivalently the Hilbert space for one site, the multiplicity $f^\alpha$ of an irrep, i.e. the number of times it appears, is given by $f^\alpha = n!/(\prod_{i=1}^n l_i)$, where the hook length $l_i$ of a box is defined as the number of boxes on the same row at the right plus the number of boxes in the same column below plus the box itself (see Fig.1). The multiplicity is equal to the number of *standard Young tableaux*, i.e. Young tableaux filled up with numbers from 1 to $n$ in ascending order from left to right in any row, and from top to bottom in any column. The standard Young tableaux can be ranked from 1 to $f^\alpha$ through the last letter sequence: two standard tableaux $S_r$ and $S_s$ are such that $S_r < S_s$ if the number $n$ appears in a row below the one it appears in $S_s$. If those rows are the same, one looks at the rows of $n - 1$, etc (see Fig.1). The dimension $d_N^\alpha$ of an irrep can also be calculated very simply from the shape $\alpha$ as $d_N^\alpha = \prod_{i=1}^n (d_{i,N}/l_i)$, where $d_{i,N} = N + \gamma_i$, where $\gamma_i$ is the algebraic distance from the $i^{th}$ box to the main diagonal, counted positively (resp. negatively) for a box above (below) the diagonal (see Fig.1). The full Hilbert space can be decomposed as $\square^{\otimes n} = \oplus_\alpha V^\alpha$, where $V^\alpha$ is the Hilbert space associated to irrep $\alpha$, and, if $d_N^\alpha > 1$, $V^\alpha$ can itself be decomposed into $d_N^\alpha$ equivalent sub-sectors $V_i^\alpha$ as $V^\alpha = \oplus_i V_i^\alpha$, with $\dim(V_i^\alpha) = f^\alpha$, $\dim(V^\alpha) = f^\alpha d_N^\alpha$ and $\dim(\square^{\otimes n}) = N^n = \sum_\alpha f^\alpha d_N^\alpha$[27].

For our purpose, the key property is that, since it has SU(N)

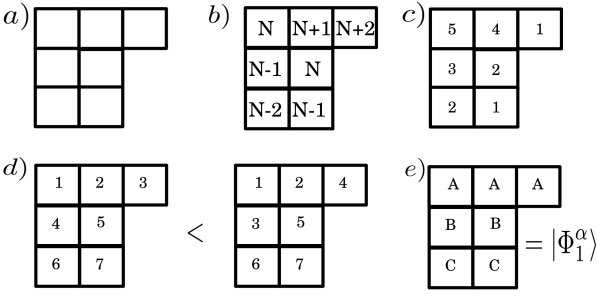

FIG. 1. a) Example of a Young tableau: $\alpha = [3,2,2]$; b) Integers $d_{i,N}$ that enter the numerator of the dimension of $\alpha$; c) Hook lengths $l_i$; d) Examples of standard tableaux ranked according to the *last letter sequence*. e) Normal product state $|\Phi_1^{[3,2,2]}\rangle = |AAABBCC\rangle$.

symmetry, the quantum permutation Hamiltonian $H$ can be diagonalized independently in each sub-sector $V_i^\alpha$, whose size (in particular that of the singlet) becomes much smaller than that of the Hilbert space used in standard ED when $N$ increases (see examples in the table of Fig.2 and Supplemental Material [28]). To diagonalize $H$ directly in a sub-sector $V_i^\alpha$, one should construct an orthonormal basis of this sector, and to write the matrix of $H$ in this basis. In principle, one can construct a basis recursively using $SU(N)$ Clebsch-Gordan coefficients[29]. However, since the multiplicity of an irrep is equal to the number of standard Young tableaux, a natural alternative is to try and associate directly a basis state to each standard Young tableau. This can be achieved by using Young symmetrization operator, the product of antisymmetrizers on the columns followed by symmetrizers on the rows [30]. Indeed, one can get a set of $f^\alpha$ linearly independent states that all belong to irrep $\alpha$ by applying the Young symmetrization operator associated to a standard tableau $S_r$, in which the sites involved in the symmetrizers and antisymmetrizers are chosen according to the numbering of $S_r$, to the product state : $|\Phi_r^\alpha\rangle = |\sigma_1\rangle \otimes ... \otimes |\sigma_n\rangle$, with $|\sigma_i\rangle = A$ if $i$ belongs to the first line of $S_r$, $B$ if it belongs to its second line, etc [28]. However, this construction does not lead to a simple method to perform ED of the $SU(N)$ Heisenberg model for two reasons. First, these states are not orthogonal. Besides, the Hamiltonian does not take a simple form.

In his substitutional analysis, Young also realized that the Young symmetrization operators (called *natural units* in his original work[31] ) were not convenient to solve algebraic problems [32]. So, he further developed the theory of the permutation group to come up with more powerful operators than the simple products of symmetrizers and antisymmetrizers . More specifically, he constructed linear superpositions of permutations of the symmetric group $S_n$ that he called *orthogonal units* which, when interpreted as operators in the Hilbert space of the $SU(N)$ Heisenberg model, will enable us to construct an orthonormal basis in which the quantum permutation Hamiltonian takes a very simple form. For a fixed shape $\alpha$, there are $(f^\alpha)^2$ orthogonal units $\{o_{rs}^\alpha\}_{r,s=1...f^\alpha}$, where the indices $r$ and $s$ refer to two standard tableaux of shape

$\alpha$. They can be constructed recursively as nested products of symmetrizers and antisymmetrizers associated to standard tableaux of smaller size [28, 33]. The resulting expressions are rather complicated. For instance, for the shape $[2,1]$ ($n = 3$), for which there are two standard tableaux, the first orthogonal unit reads $o_{11}^{[2,1]} = \frac{1}{12}(\epsilon + \tau_{1,2})^2(\epsilon - \tau_{1,3})(\epsilon + \tau_{1,2})$ where $\tau_{i,j}$ denotes the transposition $i \leftrightarrow j$ and $\epsilon$ is the identity for $S_n$. In practice however, we will never need the explicit expressions of the orthogonal units, but only some of their properties derived by Young that we now summarize.

(i) They satisfy orthonormal relations: $\forall \alpha, \beta$

$$o_{rs}^\alpha o_{uv}^\beta = \delta^{\alpha\beta} \delta_{su} o_{rv}^\alpha \quad \forall\, r,s = 1...f^\alpha, \forall\, u,v = 1...f^\beta. \quad (2)$$

(ii) The projector onto the irrep $\alpha$ can be decomposed as $T^\alpha = \sum_{r=1...f^\alpha} o_{rr}^\alpha$.

(iii) They provide a basis in which every linear superposition of permutations of $S_n$ can be uniquely decomposed, a simple consequence of Eq.(2) and of the identity $\sum_\alpha f_\alpha^2 = n!$ [31]. In particular, the Hamiltonian of Eq.(1) can be written as $H = \sum_{\beta,t,q} \mu_{tq}^\beta(H) o_{tq}^\beta$, where $\mu_{tq}^\beta(H)$ are real coefficients.

(iv) The decomposition of *successive transpositions*, i.e. transpositions between consecutive numbers $\tau_{k,k+1}$ ($1 \leq k \leq n-1$), takes a very simple form. In fact, if we write $\tau_{k,k+1} = \sum_{\beta,t,q} \mu_{tq}^\beta(\tau_{k,k+1}) o_{tq}^\beta$, then, for a given shape $\alpha$, the matrices $\bar{\mu}^\alpha(\tau_{k,k+1})$ defined by $[\bar{\mu}^\alpha(\tau_{k,k+1})]_{tu} = \mu_{tu}^\alpha(\tau_{k,k+1})$ are orthogonal and very sparse, with at most two non-vanishing entries per column or per line that can be calculated easily. More precisely, if $k+1$ and $k$ are in the same row (resp. column) in $S_t$, then $\mu_{tt}^\alpha(\tau_{k,k+1}) = +1$ (resp. $-1$), and all other matrix elements involving $t$ vanish. If $k+1$ and $k$ are not in the same column or the same line, and if $S_u$ is the tableau obtained from $S_t$ by interchanging $k$ and $k+1$, then the only non-vanishing matrix elements involving $t$ or $u$ are given by

$$\begin{pmatrix} \mu_{tt}^\alpha(\tau_{k,k+1}) & \mu_{tu}^\alpha(\tau_{k,k+1}) \\ \mu_{ut}^\alpha(\tau_{k,k+1}) & \mu_{uu}^\alpha(\tau_{k,k+1}) \end{pmatrix} = \begin{pmatrix} -\rho & \sqrt{1-\rho^2} \\ \sqrt{1-\rho^2} & \rho \end{pmatrix}$$

where $\rho$ is the inverse of *the axial distance* from $k$ to $k+1$ in $S_t$ defined by counting $+1$ (resp. $-1$) for each step made downwards or to the left (resp. upwards or to the right) to reach $k+1$ from $k$. For instance, $\tau_{3,4}$ has non vanishing matrix elements between the two tableaux of Fig.1 d), with diagonal matrix elements equal to $-1/3$ (left tableau) and $1/3$ (right tableau), and off-diagonal matrix elements equal to $2\sqrt{2}/3$.

(v) The matrix $\bar{\mu}^\beta(\sigma)$ that enters the decomposition of any permutation $\sigma = \sum_{\beta,t,q} \mu_{tq}^\beta(\sigma) o_{tq}^\beta$ is also orthogonal. Indeed, a permutation can be decomposed as a product of transpositions, and any transposition $\tau_{i,j}$ can be decomposed as a product of successive transpositions according to (assuming $i < j$):

$$\tau_{i,j} = \tau_{i,i+1}\tau_{i+1,i+2}...\tau_{j-1,j}\tau_{j-2,j-1}...\tau_{i+1,i+2}\tau_{i,i+1},$$

so that the matrix $\bar{\mu}^\beta(\sigma)$ is a product of orthogonal matrices.

To prove the central results of this paper, we need an additional property not derived by Young:

**Lemma:** When interpreted as operators acting in the Hilbert space, the orthogonal units satisfy:

$$(o_{rs}^\beta)^\dagger = o_{sr}^\beta \qquad (3)$$

*Proof:* The decomposition of the permutations can be inverted as $o_{rs}^\beta = \frac{f^\alpha}{n!} \sum_{\sigma \in S_n} \mu_{sr}^\beta(\sigma^{-1})\sigma$[31, 32]. Now, $\bar{\bar{\mu}}^\beta(\sigma^{-1}) = [\bar{\bar{\mu}}^\beta(\sigma)]^{-1}$, and since $\bar{\bar{\mu}}^\beta(\sigma)$ is orthogonal, $[\bar{\bar{\mu}}^\beta(\sigma)]_{sr}^{-1} = [\bar{\bar{\mu}}^\beta(\sigma)]_{rs}$, so that $\mu_{sr}^\beta(\sigma^{-1}) = \mu_{rs}^\beta(\sigma)$. Then, since the adjoint operator of any permutation $\sigma$ is $\sigma^\dagger = \sigma^{-1}$, we can write:

$$(o_{rs}^\beta)^\dagger = \frac{f^\alpha}{n!}\sum_{\sigma \in S_n}\mu_{sr}^\beta(\sigma^{-1})\sigma^\dagger = \frac{f^\alpha}{n!}\sum_{\sigma \in S_n}\mu_{rs}^\beta(\sigma)\sigma^{-1} = o_{sr}^\beta.$$

We are now in a position to state and demonstrate the two central results of this paper.

**Proposition 1:** Let $|\Phi_1^\alpha\rangle$ be the product state associated to the first standard tableau $S_1$ (see Fig.1e). Then, the set

$$\left\{ |\Psi_r^\alpha\rangle = ||o_{11}^\alpha|\Phi_1^\alpha\rangle||^{-1}o_{r1}^\alpha|\Phi_1^\alpha\rangle \right\}_{r=1\dots f^\alpha}$$

is an orthonomal basis of one of the sub-sectors of $V^\alpha$.

*Proof:* $\langle \Psi_r^\alpha|\Psi_r^\alpha\rangle = ||o_{11}^\alpha|\Phi_1^\alpha\rangle||^{-2}\langle\Phi_1^\alpha|(o_{r1}^\alpha)^\dagger o_{r1}^\alpha|\Phi_1^\alpha\rangle = ||o_{11}^\alpha|\Phi_1^\alpha\rangle||^{-2}\langle\Phi_1^\alpha|o_{1r}^\alpha o_{r1}^\alpha|\Phi_1^\alpha\rangle = \delta_{lr}$, where we have used $o_{11}^\alpha|\Phi_1^\alpha\rangle \neq 0$ [28]. Besides, the states transform according to the irrep $\alpha$ since $o_{r1}^\alpha|\Phi_1^\alpha\rangle = o_{rr}^\alpha o_{r1}^\alpha|\Phi_1^\alpha\rangle = T^\alpha o_{r1}^\alpha|\Phi_1^\alpha\rangle \in V^\alpha$. Finally, from Eq.(2), the set is obviously invariant under any permutation. So it must generate a sub-sector of $V^\alpha$.

**Proposition 2:** The matrix elements of $H$ in this basis are

$$\langle\Psi_r^\alpha|H|\Psi_s^\alpha\rangle = \mu_{rs}^\alpha(H). \qquad (4)$$

*Proof:* This is a simple consequence of Eq.(2).

These coefficients are simply related to those of transpositions by $\mu_{rs}^\beta(H) = \sum_{(i,j)} J_{ij}\mu_{rs}^\beta(\tau_{i,j})$, which are themselves products of the sparse matrices of successive transpositions whose explicit form has been given in point iv) above. So, we have succeeded in constructing an orthonormal basis of one sub-sector of any irrep, and we have come up with a very simple scheme to construct the matrix of the Hamiltonian in this basis. Let us emphasize that the explicit calculation of the basis is *not* required to calculate the matrix elements of the Hamiltonian or of any operator that can be written as a permutation.

We have used this theory to numerically investigate the antiferromagnetic Heisenberg $SU(N)$ Hamiltonian on the square lattice (Eq.(1) with $J_{ij} = J$ for pairs of nearest neighbors and 0 otherwise) for $N = 5$ (20 and 25 sites), $N = 8$ (16 sites), and $N = 10$ (20 sites). In each case, we have calculated the real-space correlations $\langle P_{0j}\rangle - 1/N$, the low energy spectrum, and the dimer-dimer correlations $\langle P_{ij}P_{\text{ref}}\rangle - \langle P_{\text{ref}}\rangle^2$. Some basic information (size of Hilbert space, ground state energy) is summarized in the table of Fig.2.

As can be seen in Fig.2, short-range color order is clearly present in all cases, with positive correlations which point to an $N$-site periodicity. To check whether these correlations are actually long-ranged, the best way with ED is to look at the low-energy spectrum, which is expected to build an Anderson tower of states[36–38] if the $SU(N)$ symmetry is broken in the ground state[9]. As can be seen from Fig.3, this is clearly the case for $SU(5)$. We have actually been able to calculate the full low-energy spectrum for $N = 5$ on 20 sites, whose structure illustrates several general features of $SU(N)$ models[28]. The ordering pattern suggested by real-space correlations is consistent with linear flavor-wave theory which predicts that, up to a mirror reflexion (and of course to color permutation), there is a single pattern (shown as an inset of the $SU(5)$ tower of states in Fig.3) able to minimize the zero point energy on each bond[28].

For $SU(8)$, the spectrum has a very different structure: there is a 3-fold degenerate singlet far below the first non-singlet excited states, as in the case of $SU(4)$, and with the same quantum numbers[14, 19]. This is typical of a translational symmetry breaking, and the quantum numbers (two states of zero momentum, one state of momentum $(\pi,0)$ and one state of momentum $(0,\pi)$) are compatible with a sponta-neous dimerization with columns or rows of dimers. This possibility is clearly confirmed by the dimer-dimer correlations of Fig.4, which point to well developed long-range dimer order. These correlations are very similar to those of $SU(4)$, in which case iPEPS simulations have been able to further confirm the nature of the instability[19]. So ED clearly point to spontaneous dimerization for $SU(8)$. Of course, since it takes eight (or a multiple of eight) sites to build a singlet for $SU(8)$, the dimers are not singlets, but they build an irrep of dimension $N(N-1)/2 = 28$. Whether these effective degrees of freedom develop some kind of order cannot be decided on the basis of the present results.

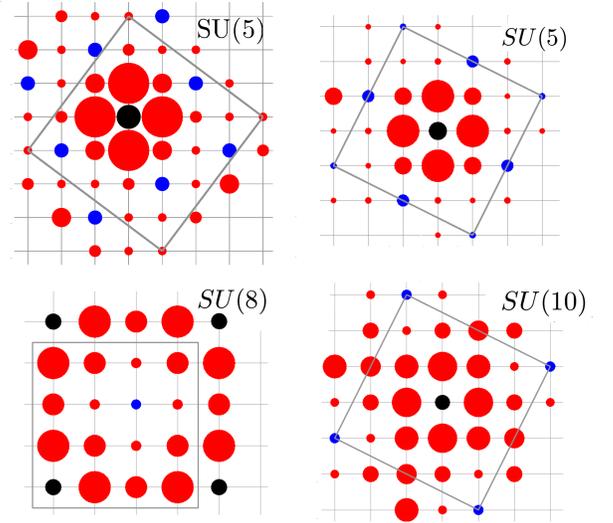

FIG. 2. . (Color online) Real-space correlations $\langle P_{0j}\rangle - 1/N$ for various $SU(N)$ models and cluster sizes on the square lattice with periodic boundary conditions: $SU(5)$ (tilted 25 and 20 site cluster), $SU(8)$ (16 sites) and $SU(10)$ (20 sites). The black dot is the reference site 0. Positive (negative) correlations are depicted as blue (red) disks with an area proportionnal to the absolute value of the correlation. The correlations for $SU(5)$ on the $(5 \times 5)$ 25-site cluster are shown and discussed in the Supplemental Material[28]. Table: dimension $f^{[\frac{n}{N},\dots,\frac{n}{N}]}$ of the singlet subspace in which the permutation Hamiltonian has been diagonalized, approximate dimension $(n-1)!/(\frac{n}{N})!^N$ of the Hilbert space used in standard ED [34], ground states energies per site $\mathcal{E}_{GS}$.

| $SU(N)$ | $n$ | $f^{[k,\dots,k]}$ | $\frac{(n-1)!}{k!^N}$ | $\mathcal{E}_{GS}$ |
|---|---|---|---|---|
| $SU(5)$ | 25 (tilted) | 701149020 | $2.5 \times 10^{13}$ | $-1.154334$ |
| $SU(5)$ | 25 ($5 \times 5$) | 701149020 | $2.5 \times 10^{13}$ | $-1.164712$ |
| $SU(5)$ | 20 | 1662804 | $1.5 \times 10^{10}$ | $-1.215377$ |
| $SU(8)$ | 16 | 1430 | $5.1 \times 10^9$ | $-1.572223$ |
| $SU(10)$ | 20 | 16796 | $1.2 \times 10^{14}$ | $-1.589218$ |

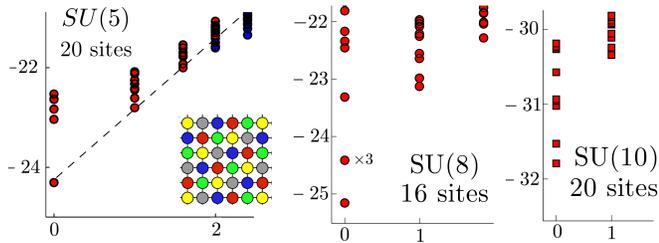 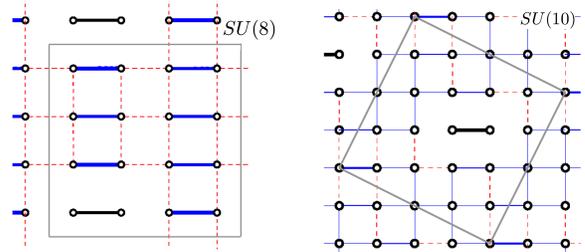

FIG. 3. Energy spectra (in units of $J$) for $SU(5)$ (20 sites), $SU(8)$ (16 sites) and $SU(10)$ (20 sites), plotted as a function of the quadratic Casimir $C_2$ (See [28, 35]). Different irreps with the same $C_2$ ( e.g. $C_2 = 2$ for $SU(5)$) are represented with different colors. Below the $SU(5)$ tower of states, sketch of the long-ranged color ordered pattern consistent with the real-space correlations of Fig. 2 and with flavor-wave theory.

FIG. 4. . Dimer-dimer correlations $\langle P_{ij}P_{\text{ref}}\rangle - \langle P_{\text{ref}}\rangle^2$ for $SU(8)$ (16 sites) and $SU(10)$ (20 sites). The reference bond is shown in black while positive (negative) correlations are shown as solid blue (dashed red) lines, with a thickness proportionnal to the dimer-dimer correlation.

Finally, the case of $SU(10)$ is again quite different. In that case, dimer-dimer correlations do not point to any kind of dimer order, and there are several low-lying singlets below the first non-singlet excitation. This is reminiscent of the situation observed in $SU(2)$ quantum spin liquids, such as the kagome antiferromagnet[39–41]. So the most likely possibility is that this system is a quantum spin liquid.

To summarize, we have introduced a simple and explicit formulation of the quantum permutation Hamiltonian separately in each irreducible representation of $SU(N)$. We have illustrated the power of the method on a problem of considerable current interest, the properties of ultra-cold multicomponent fermionic atoms loaded in an optical lattice, opening the way to the investigation of much larger values of $N$ than accessible so far. This approach is also expected to be very powerful on other problems. For instance, it should be competitive even for smaller values of $N$ in the presence of disorder since there is no spatial symmetry to reduce the size of the Hilbert of standard exact diagonalizations. The method can also be extended to the general case of the $SU(N)$ Heisenberg model with any irrep at each site, a model relevant e.g. to Mott phases with more than one fermion per site. Work is in progress along these lines.

The authors acknowledge P. Corboz, M. Hermele, A. Läuchli, K. Penc, N. Schellekens and J-B. Zuber for useful discussions. This work has been supported by the Swiss National Fund.

[1] G. Pagano, M. Mancini, G. Cappellini, P. Lombardi, F. Schafer, H. Hu, X.-J. Liu, J. Catani, C. Sias, M. Inguscio, and L. Fallani, Nat Phys **10**, 198 (2014).

[2] F. Scazza, C. Hofrichter, M. Höfer, P. C. De Groot, I. Bloch, and S. Fölling, ArXiv e-prints (2014), arXiv:1403.4761 [cond-mat.quant-gas].

[3] X. Zhang, M. Bishof, S. L. Bromley, C. V. Kraus, M. S. Safronova, P. Zoller, A. M. Rey, and J. Ye, ArXiv e-prints (2014), arXiv:1403.2964 [cond-mat.quant-gas].

[4] C. Wu, J.-p. Hu, and S.-c. Zhang, Phys. Rev. Lett. **91**, 186402 (2003).

[5] C. Honerkamp and W. Hofstetter, Phys. Rev. Lett. **92**, 170403 (2004).

[6] M. A. Cazalilla, A. F. Ho, and M. Ueda, New Journal of Physics **11**, 103033 (2009).

[7] A. V. Gorshkov, M. Hermele, V. Gurarie, C. Xu, P. S. Julienne, J. Ye, P. Zoller, E. Demler, M. D. Lukin, and A. M. Rey, Nat Phys **6**, 289 (2010).

[8] N. Papanicolaou, Nuclear Physics B **305**, 367 (1988).

[9] T. A. Tóth, A. M. Läuchli, F. Mila, and K. Penc, Phys. Rev. Lett. **105**, 265301 (2010).

[10] B. Bauer, P. Corboz, A. M. Läuchli, L. Messio, K. Penc, M. Troyer, and F. Mila, Phys. Rev. B **85**, 125116 (2012).

[11] K. I. Kugel' and D. I. Khomskiĭ, Soviet Physics Uspekhi **25**, 231 (1982).

[12] Y. Q. Li, M. Ma, D. N. Shi, and F. C. Zhang, Phys. Rev. Lett. **81**, 3527 (1998).

[13] M. van den Bossche, P. Azaria, P. Lecheminant, and F. Mila, Phys. Rev. Lett. **86**, 4124 (2001).

[14] M. van den Bossche, F.-C. Zhang, and F. Mila, The European Physical Journal B - Condensed Matter and Complex Systems **17**, 367 (2000).

[15] B. Sutherland, Phys. Rev. B **12**, 3795 (1975).

[16] B. Frischmuth, F. Mila, and M. Troyer, Phys. Rev. Lett. **82**, 835 (1999).

[17] L. Messio and F. Mila, Phys. Rev. Lett. **109**, 205306 (2012).

[18] A. Joshi, M. Ma, F. Mila, D. N. Shi, and F. C. Zhang, Phys. Rev. B **60**, 6584 (1999).

[19] P. Corboz, A. M. Läuchli, K. Penc, M. Troyer, and F. Mila, Phys. Rev. Lett. **107**, 215301 (2011).

[20] P. Corboz, M. Lajkó, A. M. Läuchli, K. Penc, and F. Mila, Phys. Rev. X **2**, 041013 (2012).

[21] P. Corboz, M. Lajkó, K. Penc, F. Mila, and A. M. Läuchli,

Phys. Rev. B **87**, 195113 (2013).

[22] F. Wang and A. Vishwanath, Phys. Rev. B **80**, 064413 (2009).

[23] P. Corboz, K. Penc, F. Mila, and A. M. Läuchli, Phys. Rev. B **86**, 041106 (2012).

[24] M. Hermele, V. Gurarie, and A. M. Rey, Phys. Rev. Lett. **103**, 135301 (2009).

[25] M. Hermele and V. Gurarie, Phys. Rev. B **84**, 174441 (2011).

[26] G. Szirmai, E. Szirmai, A. Zamora, and M. Lewenstein, Phys. Rev. A **84**, 011611 (2011).

[27] H. Weyl, *The Theory of Groups and Quantum Mechanics* (Dover Publication, 1931, 1931).

[28] See supplemental material for some definitions and proofs, as well as for additional numerical results on $SU(5)$.

[29] A. Alex, M. Kalus, A. Huckleberry, and J. von Delft, Journal of Mathematical Physics **52**, 023507 (2011).

[30] J. F. Cornwell, *Group Theory in Physics* (Academic Press, 1984).

[31] A. Young, Proc. London. Math. Soc. **2**, 196 (1931).

[32] D. E. Rutherford, *Substitutional Analysis* (Edinburgh University Press, 1948).

[33] R. M. Thrall, Duke Math. J. **8** (1941).

[34] A. M. Läuchli, *Introduction to Frustrated Magnetism*, edited by C. Lacroix, P. Mendels, and F. Mila, Springer Series in Solid-State Sciences (Springer Berlin Heidelberg, 2011) pp. 481–511.

[35] K. Pilch and A. N. Schellekens, Journal of Mathematical Physics **25** (1984).

[36] P. W. Anderson, Phys. Rev. **86**, 694 (1952).

[37] B. Bernu, C. Lhuillier, and L. Pierre, Phys. Rev. Lett. **69**, 2590 (1992).

[38] B. Bernu, P. Lecheminant, C. Lhuillier, and L. Pierre, Phys. Rev. B **50**, 10048 (1994).

[39] C. Waldtmann, H.-U. Everts, B. Bernu, C. Lhuillier, P. Sindzingre, P. Lecheminant, and L. Pierre, The European Physical Journal B - Condensed Matter and Complex Systems **2**, 501 (1998).

[40] F. Mila, Phys. Rev. Lett. **81**, 2356 (1998).

[41] M. Mambrini and F. Mila, The European Physical Journal B - Condensed Matter and Complex Systems **17**, 651 (2000).

# Supplementary material to "Exact Diagonalization of Heisenberg $SU(N)$ models".

Pierre Nataf[,*] and Frederic Mila[,†]
(Dated: July 18, 2014)

In this supplemental material, we first provide definitions and simple examples of the *orthogonal* units $\{o_{rs}^{\alpha}\}_{\alpha,r,s}$. Then, we introduce the *natural* units $\{g_{rs}^{\alpha}\}_{\alpha,r,s}$ in order to prove several properties. In particular, for a given shape $\alpha = [\alpha_1, \alpha_2, ..., \alpha_k]$ ($k \leq N$), we show that $o_{11}^{\alpha}|\Phi_1^{\alpha}\rangle \neq 0$ where $o_{11}^{\alpha}$ is the diagonal orthogonal unit of the *first* standard Young tableau of shape $\alpha$ and $|\Phi_1^{\alpha}\rangle$ is the product state associated to the first standard tableau. We also prove the linear independence of the family of states $\{g_{rr}^{\alpha}|\Phi_r^{\alpha}\rangle\}_{r=1..f^{\alpha}}$, where the state $|\Phi_r^{\alpha}\rangle$ is the product state associated to the $r^{th}$ standard tableau. In the next section, we derive the asymptotic behavior for large $N$ of the ratio of the dimension of the sector usually considered in standard ED to the singlet subspace. Then, we discuss the order and analyze the correlations of the $SU(5)$ model on the $(5 \times 5)$ 25-site cluster. Finally, we show and discuss the full low energy spectra of the $SU(5)$ model on the 20-site cluster.

## DEFINITION OF THE ORTHOGONAL UNITS $\{o_{rs}^{\alpha}\}_{\alpha,r,s}$

The orthogonal units $o_{rs}^{\alpha}$, first introduced by Young[1], can be calculated [2] at the end of a sequence involving substitutional expressions $o_{r[l]}^{\alpha[l]}$ ($1 \leq l < n$) relative to sub-standard tableaux of $l$ boxes $S_{r[l]}$, of shape $\alpha[l]$, obtained from $S_r$ by keeping the first $l$ numbers and removing the last $n-l$ numbers. This sequence is defined by, $\forall r, s \in [\![1; f^{\alpha}]\!]$:

$$
\begin{cases}
o_{r[1]}^{\alpha[1]} = \epsilon, & (\epsilon \text{ being the identity of } \mathcal{S}_n) \\[2mm]
o_{r[l]}^{\alpha[l]} = \frac{f^{\alpha[l]}}{l!} o_{r[l-1]}^{\alpha[l-1]} E_{r[l]r[l]}^{\alpha[l]} o_{r[l-1]}^{\alpha[l-1]} & \forall 1 < l < n, \\[2mm]
o_{rs}^{\alpha} = \frac{f^{\alpha}}{n!} \sqrt{\frac{\Theta_r^{\alpha}}{\Theta_s^{\alpha}}} o_{r[n-1]}^{\alpha[n-1]} E_{rs}^{\alpha} o_{s[n-1]}^{\alpha[n-1]},
\end{cases}
\tag{1}
$$

where $E_{tu}^{\gamma} = \mathcal{P}_t^{\gamma} \mathcal{N}_t^{\gamma} \sigma_{tu}^{\gamma}$. In this expression, $\mathcal{P}_t^{\gamma}$ (resp. $\mathcal{N}_t^{\gamma}$) is the sum of all the elements of the product of the positive (resp. negative) symmetric group of the rows (resp. columns) of the standard tableau $S_t$, also called the Young symmetrizer (reps. Young antisymmetrizer) of the standard tableau $S_t$, while $\sigma_{tu}^{\gamma}$ is the permutation that transforms $S_u$ into $S_t$, two standard tableaux of shape $\gamma$. Finally, $\Theta_r^{\alpha} = \phi_{S_r} \Pi_{l=1}^{n-1} \phi_{S_{r[l]}}$, with $\phi_{S_{r[l]}} = \Pi_{\nu}'(1 + c_{\nu})$, where $1/c_{\nu}$ is the *axial distance* from the last number of the row $\nu$ to $l$. The product $\Pi_{\nu}'$ it limited to the rows $\nu$ which, in $S_{r[l]}$, lie above the row containing the last number $l$.

For the example of the standard tableau $S_1 = \begin{array}{|c|c|}\hline 1 & 2 \\\hline 3 \\\cline{1-1}\end{array}$, the negative symmetric group of the second column is reduced to $\{\epsilon\}$, and that of the first column is $\{\epsilon, -\tau_{1,3}\}$, which corresponds to the set of all the permutations that mix the numbers of the considered column with their respective signature. Then, $\mathcal{N}_1^{\alpha} = (\epsilon - \tau_{1,3})$. The positive symmetric group of the first row is $\{\epsilon, \tau_{1,2}\}$, and that of the second row is reduced to $\{\epsilon\}$, so that $\mathcal{P}_1^{\alpha} = (\epsilon + \tau_{1,2})$.

Thus, for the example of the shape $[2,1]$ ($n = 3$), $S_1 = \begin{array}{|c|c|}\hline 1 & 2 \\\hline 3 \\\cline{1-1}\end{array} < S_2 = \begin{array}{|c|c|}\hline 1 & 3 \\\hline 2 \\\cline{1-1}\end{array}$, $S_{2[2]} = \begin{array}{|c|}\hline 1 \\\hline 2 \\\hline\end{array}$ and $S_{1[2]} = \begin{array}{|c|c|}\hline 1 & 2 \\\hline\end{array}$, leading to $o_{11}^{[2,1]} = \frac{1}{12}(\epsilon + \tau_{1,2})^2 (\epsilon - \tau_{1,3})(\epsilon + \tau_{1,2}) = \frac{1}{6}(2\epsilon + 2\tau_{1,2} - \tau_{1,3} - \tau_{2,3} - \tau_{1,3}\tau_{1,2} - \tau_{1,2}\tau_{1,3})$ and $o_{21}^{[2,1]} = \frac{1}{6\sqrt{3}}(\epsilon - \tau_{1,2})(\epsilon + \tau_{1,3})(\epsilon - \tau_{1,2})\tau_{2,3}(\epsilon + \tau_{1,2}) = \frac{1}{2\sqrt{3}}(\tau_{2,3} - \tau_{1,3} + \tau_{1,2}\tau_{13} - \tau_{1,3}\tau_{1,2})$, where $\tau_{i,j}$ is the permutation $i \leftrightarrow j$. Then, the normal product state $|\Phi_1^{2,1}\rangle = |AAB\rangle$ (color $A$ on sites $1, 2$, color $B$ on site 3), and the basis $\left\{ |\Psi_r^{[2,1]}\rangle = ||o_{11}^{2,1}|\Phi_1^{2,1}\rangle||^{-1} o_{r1}^{2,1}|\Phi_1^{2,1}\rangle \right\}_{r=1,2}$ is given by:

$$
|\Psi_1^{[2,1]}\rangle = \frac{1}{\sqrt{6}}(2|AAB\rangle - |BAA\rangle - |ABA\rangle)
\tag{2}
$$

$$
|\Psi_2^{[2,1]}\rangle = \frac{1}{\sqrt{2}}(-|BAA\rangle + |ABA\rangle)
\tag{3}
$$

As expected, this basis is orthonormal.

## YOUNG SYMMETRIZATION OPERATOR AND NATURAL UNITS $\{g_{rs}^\alpha\}_{\alpha,r,s}$

The Young symmetrization operators, namely the products of anitsymmetrizers on the columns of a standard tableau followed by products of symmetrizers on the lines, are the diagonal members of a more general family $\{g_{rs}^\alpha\}_{\alpha,r,s}$ first introduced by Young that he called natural units. In this section, we define them and discuss some of their properties mentioned in the main text. The diagonal ones will also be directly useful in proving the important property of the orthogonal units of next section.

The diagonal natural units $g_{rr}^\alpha$ are simply defined by

$$g_{rr}^\alpha = \frac{f^\alpha}{n!} \mathcal{P}_r^\alpha \mathcal{N}_r^\alpha \qquad \forall r = 1 .... f^\alpha, \tag{4}$$

where $\mathcal{P}_r^\alpha$ and $\mathcal{N}_r^\alpha$ were defined in the previous section.

The definition of the off-diagonal ones $g_{rs}^\alpha$ is more involved:

$$g_{rs}^\alpha = \frac{f^\alpha}{n!} P_r^\alpha N_r^\alpha \left\{ \sum_{t=1...f^\alpha} \eta_{ts}^\alpha \sigma_{rt}^\alpha \right\} \qquad \forall r \neq s \in \{1....f^\alpha\}, \tag{5}$$

where $\sigma_{rt}^\alpha$ is the permutation which changes the standard tableau $t$ into the standard tableau $r$ , and where $\eta_{ts}^\alpha$ is the $t, s$ element of the inverse of the matrix $\Xi^\alpha$. $\Xi^\alpha$ is a $f^\alpha \times f^\alpha$ matrix whose $q, u$ element is equal to the coefficient of $\epsilon$ in the product $\mathcal{N}_u \mathcal{P}_q$.

Their properties are very similar to the orthogonal units $o_{rs}^\alpha$; they also satisfy [3]:

$$g_{rs}^\alpha g_{uv}^\beta = \delta^{\alpha\beta} \delta_{su} g_{rv}^\alpha \tag{6}$$

$$\epsilon = \sum_{\alpha, r} g_{rr}^\alpha. \tag{7}$$

The *projection operator* $T^\alpha$, introduced in the main text, can also be written as a sum over the $f^\alpha$ diagonal natural $g_{rr}^\alpha$ units, so that:

$$T^\alpha = \sum_r g_{rr}^\alpha = \sum_r o_{rr}^\alpha. \tag{8}$$

The definition of the natural units is much simpler than that of the orthogonal units, and the reader might wonder why we did not use this simpler representation. There are good reasons for that. First of all, the coordinates $\lambda_{rs}^\alpha(H)$ which appear in the decomposition of the quantum permutation Hamiltonian $H = \sum_{\alpha,r,s} \lambda_{rs}^\alpha(H) g_{rs}^\alpha$ are not as simple as for the orthogonal units $o_{rs}^\alpha$. Besides, and more importantly, by contrast to the orthogonal units, $(g_{rs}^\alpha)^\dagger \neq g_{sr}^\alpha$, so that there is no simple way to construct an orthonormal basis with their help. For instance, if we look at the example of the shape $[2,1]$, the non-diagonal natural unit $g_{21}^{[2,1]}$ reads:

$$g_{21}^{[2,1]} = \frac{1}{3} \mathcal{P}_2^{[2,1]} \mathcal{N}_2^{[2,1]} (\tau_{2,3}) = \frac{1}{3} (\epsilon + \tau_{1,3})(\epsilon - \tau_{1,2})(\tau_{2,3}) = \frac{1}{3} \{\tau_{2,3} - \tau_{1,2} - \tau_{1,2}\tau_{2,3} + \tau_{1,3}\tau_{2,3}\}, \tag{9}$$

while:

$$g_{12}^{[2,1]} = \frac{1}{3} \mathcal{P}_1^{[2,1]} \mathcal{N}_1^{[2,1]} (\tau_{2,3}) = \frac{1}{3} (\epsilon + \tau_{1,2})(\epsilon - \tau_{1,3})(\tau_{2,3}) = \frac{1}{3} \{\tau_{2,3} - \tau_{1,3} + \tau_{1,2}\tau_{2,3} - \tau_{1,3}\tau_{2,3}\}. \tag{10}$$

Thus, clearly $(g_{21}^{[2,1]})^\dagger \neq g_{12}^{[2,1]}$. And if we consider the states that we can build from the natural units $g_{11}^{[2,1]}$ , $g_{21}^{[2,1]}$ and from the normal product state $|\Phi_1^{2,1}\rangle = |AAB\rangle$ (following the construction performed with the orthogonal units), since $g_{11}^{[2,1]} = (1/3)(\epsilon + \tau_{1,2})(\epsilon - \tau_{1,3})$, one has:

$$g_{11}^{[2,1]} |\Phi_1^{[2,1]}\rangle = \frac{1}{3} \{2|AAB\rangle - |BAA\rangle - |ABA\rangle\} \tag{11}$$

$$g_{21}^{[2,1]} |\Phi_1^{[2,1]}\rangle = \frac{1}{3} \{2|ABA\rangle - |BAA\rangle - |AAB\rangle\}, \tag{12}$$

which are not orthogonal. Note that $g_{11}^{[2,1]} |\Phi_1^{[2,1]}\rangle$ is proportional to $|\Psi_1^{[2,1]}\rangle$ (defined in Eq. 2), a property that we will be able to prove below for every shape $\alpha$. This will allow us in particular to demonstrate that $o_{11}^\alpha |\Phi_1^\alpha\rangle \neq 0$.

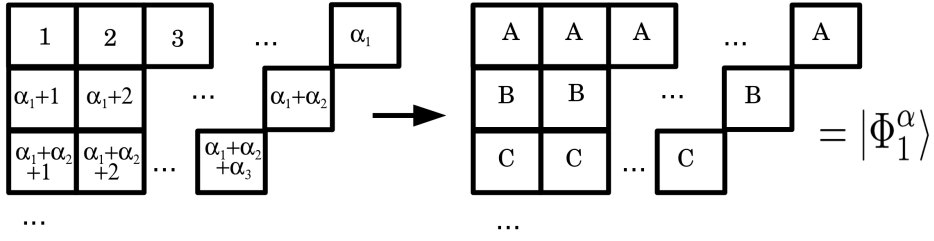

FIG. 1: From a given shape $\alpha = [\alpha_1, \alpha_2, ..., \alpha_k]$, the *first* standard Young tableau $S_1$ is filled up with the sequence $1, 2, 3, ..., \alpha_1$ in the first line, $\alpha_1 + 1, \alpha_1 + 2, ..., \alpha_1 + \alpha_2$ in the second line, etc....It is shown on the left. The state $|\Phi_1^\alpha\rangle$, shown on the right, is a product state with color $A$ on sites $1, 2, 3, ..., \alpha_1$, color $B$ on sites $\alpha_1 + 1, \alpha_1 + 2, ..., \alpha_1 + \alpha_2$, color $C$ on sites $\alpha_1 + \alpha_2 + 1$, etc...

### PROOF THAT $o_{11}^\alpha |\Phi_1^\alpha\rangle \neq 0$

**Proposition 1:** $\forall \alpha = [\alpha_1, \alpha_2, ..., \alpha_k]$ with $k \leq N$, $\langle\Phi_1^\alpha|\mathcal{P}_1^\alpha = \alpha_1!\alpha_2!...\alpha_k!\langle\Phi_1^\alpha|$, where $|\Phi_1^\alpha\rangle$ is defined in Fig. 1).

*Proof:* $\mathcal{P}_1^\alpha$ is a product of $k$ terms. The first one is the sum of the $\alpha_1!$ permutations $\sigma \in \mathcal{S}_{\alpha_1}$ which send $1, 2, ..., \alpha_1$ onto $\sigma(1), \sigma(2), ..., \sigma(\alpha_1)$, while the $q^{th}$ term (for $1 < q \leq k$) is the sum of the $\alpha_q!$ permutations $\sigma \in \mathcal{S}_{\alpha_q}$ which send the numbers $\alpha_1 + \alpha_2 + ... + \alpha_{q-1} + 1, ..., \alpha_1 + \alpha_2 + ... + \alpha_q$ onto $\sigma(\alpha_1 + \alpha_2 + ... + \alpha_{q-1} + 1), ..., \sigma(\alpha_1 + \alpha_2 + ... + \alpha_q)$. Then, since the adjoint operator of any permutation $\sigma$ is $\sigma^\dagger = \sigma^{-1}$ (as can be easily checked on the product states basis of the full Hilbert space $\{\Pi_{i=1...n}|\chi_i\rangle, \chi_i \in \{A, B, C, D...\} \, \forall i = 1...n\}$), $(\mathcal{P}_1^\alpha)^\dagger = \mathcal{P}_1^\alpha$. Consequently, $\langle\Phi_1^\alpha|\mathcal{P}_1^\alpha = \langle(\mathcal{P}_1^\alpha)^\dagger\Phi_1^\alpha| = \langle\mathcal{P}_1^\alpha\Phi_1^\alpha| = \alpha_1!\alpha_2!...\alpha_k!\langle\Phi_1^\alpha|$.

**Proposition 2:** $\langle\Phi_1^\alpha|g_{11}^\alpha|\Phi_1^\alpha\rangle > 0$

*Proof:* The antisymmetrizer $\mathcal{N}_1^\alpha$ associated with the numbers belonging to the columns of the first standard Young tableau will generate, when acting on $|\Phi_1^\alpha\rangle$, a linear superposition involving a lot of product states with $\alpha_1$ colors $A$, $\alpha_2$ colors $B$, etc.... The only product state of this superposition equal to $|\Phi_1^\alpha\rangle$ will come from the product of the identity operator for each column, and the corresponding coefficient will be $+1$. The other permutations of $\mathcal{N}_1^\alpha$ will systematically send at least one letter onto a line where the sites have different letters in $|\Phi_1^\alpha\rangle$. For instance, the transposition $\tau_{1,\alpha_1+1}$, will send a letter $A$ onto the site $\alpha_1 + 1$, and the letter $B$ onto the site 1. Consequently, all those states will have vanishing overlap with $\langle\Phi_1^\alpha|$, and $\langle\Phi_1^\alpha|\mathcal{N}_1^\alpha|\Phi_1^\alpha\rangle = 1$. Finally,

$$\langle\Phi_1^\alpha|g_{11}^\alpha|\Phi_1^\alpha\rangle = \frac{f^\alpha}{n!}\langle\Phi_1^\alpha|\mathcal{P}_1^\alpha\mathcal{N}_1^\alpha|\Phi_1^\alpha\rangle = \frac{f^\alpha\alpha_1!\alpha_2!...\alpha_k!}{n!}\langle\Phi_1^\alpha|\mathcal{N}_1^\alpha|\Phi_1^\alpha\rangle = \frac{f^\alpha\alpha_1!\alpha_2!...\alpha_k!}{n!} > 0. \tag{13}$$

**Proposition 3:** $g_{rr}^\alpha|\Phi_1^\alpha\rangle = o_{rr}^\alpha|\Phi_1^\alpha\rangle = 0$ for $r > 1$.

*Proof:* Let us denote by $j_0$ the smallest integer between 1 and $n$ which does not have in the standard tableau $S_r$ the same location as in the *first* standard Young tableau. Necessarily, $j_0$ appears in the tableau $S_r$ in the first column (otherwise the tableau would not be standard), and in a line that is below the one where $j_0$ appears in the first standard tableau. If we call in general $y_{S_q}(j)$ the line where the number $j$ appears in the standard tableau $S_q$ (for $q = 1...f^\alpha$), we have $y_{S_r}(j_0) = y_{S_1}(j_0) + 1$. Then, the antisymmetrizer of the first column of the standard tableau $S_r$ will mix $1, \alpha_1 + 1, ..., j_0, ...$etc ($k$ numbers in total). Now, an antisymmetrizer containing $k$ sites which acts on a product states which have strictly less than $k$ different letters on those $k$ sites gives 0. Moreover, among all the sites from 1 to $j_0$, there are only $y_{S_1}(j_0)$ different letters in $|\Phi_1^\alpha\rangle$, which implies that there are at least two equal letters on the sites belonging to the first column of the standard tableau $S_r$. We conclude that $\mathcal{N}_r^\alpha|\Phi_1^\alpha\rangle = 0$, and so that $g_{rr}^\alpha|\Phi_1^\alpha\rangle = 0$.

Finally, if we look at the definition of the orthogonal units $o_{rr}^\alpha$, one can write $o_{rr}^\alpha = \mathcal{A}\mathcal{N}_{r[j_0]}^{\alpha[j_0]}\mathcal{B}$, where $\mathcal{B}$ is a substitutional expression containing the symmetrizers and antisymmetrizers of the first $j_0-1$ reduced Young tableaux, i.e $\mathcal{P}_{r[j]}^{\alpha[j]}$ and $\mathcal{N}_{r[j]}^{\alpha[j]}$, with $j < j_0$. The important thing is that they do not involve numbers larger than $j_0$, so that $\mathcal{B}|\Phi_1^\alpha\rangle$ has no more than $y_{S_1}(j_0)$ different letters on sites $1, 2, ..., j_0$. Thus, since $\mathcal{N}_{r[j_0]}^{\alpha[j_0]}$ contains an antisymmetrizer with $y_{S_1}(j_0) + 1$ sites among $1, 2, ..., j_0$, it gives 0 when it acts on $\mathcal{B}|\Phi_1^\alpha\rangle$).

**Proposition 4:** $o_{11}^\alpha|\Phi_1^\alpha\rangle \neq 0$

*Proof:* Eq.8 and the results of the previous section allows one to write:

$$T^\alpha|\Phi_1^\alpha\rangle = o_{11}^\alpha|\Phi_1^\alpha\rangle = g_{11}^\alpha|\Phi_1^\alpha\rangle \neq 0$$

.

## LINEAR INDEPENDENCE OF THE FAMILY OF STATES $\{g_{rr}^\alpha|\Phi_r^\alpha\rangle\}_{r=1..f^\alpha}$.

Given a standard tableau $S_r$, one defines the product state $|\Phi_r^\alpha\rangle = |\sigma_1\rangle \otimes ... \otimes |\sigma_n\rangle$ with $|\sigma_i\rangle = A$ if $i$ belongs to the first line of $S_r$, $B$ if it belongs to the second line, etc. First, one has $g_{rr}^\alpha|\Phi_r^\alpha\rangle \neq 0 \; \forall r = 1..f^\alpha$. One can prove this using exactly the same reasoning as in propositions 1 and 2 of the previous section. Secondly, if $\sum_q \eta_q g_{qq}^\alpha|\Phi_q^\alpha\rangle = 0$, the mutliplication on the left by $g_{rr}^\alpha$ implies $\eta_r = 0 \; \forall r = 1..f^\alpha$ as a consequence of Eq. (6), which proves the linear independance of the states $\{g_{rr}^\alpha|\Phi_r^\alpha\rangle\}_{r=1..f^\alpha}$

## RELATIVE SIZE OF THE SUBSPACE SPANNED BY THE STATES IN WHICH EACH COLOR APPEARS $k$ TIMES TO THE SINGLET SUBSPACE

In the main text, we wrote that the reduction in the Hilbert space size achieved by working in the singlet sector scales like $N^n$. Since this is the dimension of the full Hilbert space, this statement requires a justification. We give it in this section by deriving the scaling form of this reduction factor in the case where the number of sites $n$ is a multiple of $N$, $n = kN$. In that case, the singlet sector $[k, ..., k]$, corresponding to a rectangular Young tableau made of $k$ columns of length $N$ has dimension $\dim(V^{[k,...,k]}) = f^{[k,...,k]}$. Now, standard ED simulations for Heisenberg $SU(N)$ models are done in the subspace of dimension $n!/k!^N$ spanned by the states in which each color appears $k$ times. We derive below the behavior of the ratio $\frac{n!/k!^N}{f^{[k,...,k]}}$ in the large $N$ limit and for fixed $k$. First of all, a direct application of the *hook length formulary* leads to:

$$f^{[k,...,k]} = \frac{n!}{k! \frac{(k+1)!}{1!} \frac{(k+2)!}{2!} \frac{(k+3)!}{3!} \cdots \frac{(k+N-1)!}{(N-1)!}}, \tag{14}$$

so that

$$\frac{\frac{n!}{k!^N}}{f^{[k,...,k]}} = \frac{k!}{k!} \frac{(k+1)!}{k!1!} \frac{(k+2)!}{k!2!} \frac{(k+3)!}{3!k!} \cdots \frac{(k+N-1)!}{k!(N-1)!} \tag{15}$$

$$= (k+1) \times \frac{(k+1)(k+2)}{2} \times \frac{(k+1)(k+2)(k+3)}{3 \times 2} \times ... \times \frac{(k+1)(k+2)...(k+N-1)}{2 \times 3 \times ... \times (N-1)}. \tag{16}$$

In the numerator of the last product, one has $N-1$ times the factor $(k+1)$, $N-2$ times the factor $(k+2)$, and more generally $N+k-j$ times the factor $j$, for all the integers $j$ between $(k+1)$ and $(k+N-1)$. The logarithm of the numerator of the last product can then be written as $\sum_{j=k+1}^{j=k+N-1}(N+k-j)\log(j) = (N+k)\log((k+N-1)!) - (N+k)\log(k!) - \sum_{j=k+1}^{j=k+N-1} j \log(j)$. In such an expression, the first term can be evaluated with the help of the Stirling's approximation:

$$\log(q!) = q\log(q) - q + \frac{1}{2}\log(2\pi q) + \frac{1}{12q} + O\left(\frac{1}{q^2}\right). \tag{17}$$

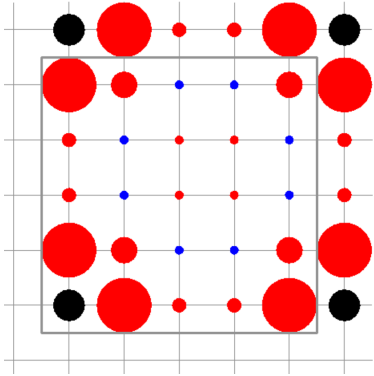

FIG. 2: Real-space correlations $\langle P_{0j} \rangle - 1/N$ for $SU(5)$ on the $5 \times 5$ 25-site cluster.

To evaluate the discrete sum $\sum_{j=k+1}^{j=k+N-1} j \log(j)$, one uses the Euler-MacLaurin formula which allows one to write:

$$\sum_{q_1}^{q_2} q \log(q) = \frac{1}{2} q_2(q_2+1) \log(q_2) - \frac{1}{2} q_1(q_1-1) \log(q_1) - \left( \frac{q_2^2}{4} - \frac{q_1^2}{4} \right) + \frac{1}{12} \log\left( \frac{q_2}{q_1} \right) + O\left( \frac{1}{q_1^3} \right) + O\left( \frac{1}{q_2^3} \right). \quad (18)$$

The same procedure can be applied to calculate the logarithm of the denominator, so after having gathered all the terms, one obtains:

$$\log\left( \frac{\frac{n!}{k!^N}}{f^{[k,\ldots,k]}} \right) = kN \log(N) - N(k + \log(k!)) + \log(N) \frac{k^2}{2} + \log C_k + \frac{1}{N} \frac{1}{6} (k^3 - \frac{k^2}{2} + k - 1) + O(1/N^2), \quad (19)$$

where the constant $C_k$ reads :

$$C_k = (k+1)^{\frac{k(k+1)}{2} + \frac{1}{12}} k!^{-k} e^{-\frac{k(k+1)}{4}} (2\pi)^{k/2} \quad (20)$$

Thus, for large $N$ ($k$ fixed), the ratio behaves as:

$$\frac{\frac{n!}{k!^N}}{f^{[k,\ldots,k]}} \simeq C_k N^n N^{\frac{k^2}{2}} \left( e(k!)^{\frac{1}{k}} \right)^{-n} e^{\frac{k^3 - \frac{k^2}{2} + k - 1}{6N}}. \quad (21)$$

The dominant behavior is $N^n$, as quoted in the paper.

## ORDERING PATTERN FOR $SU(5)$ AND CORRELATIONS ON THE ($5 \times 5$) 25-SITE CLUSTER.

According to linear flavor-wave theory, the minimum of the zero-point energy is reached as soon as each pair of neighboring colors is only coupled to colors different from both colors of the pair[5]. For $SU(5)$ on the square lattice, this is a strong constraint which, up to color permutation, leads to only two patterns related by a mirror reflexion. These patterns have a 5-site unit cell with unit vectors $\{(2,1),(-1,2)\}$ and $\{(1,2),(-2,1)\}$ respectively. The first one is depicted as an inset of the $SU(5)$ tower of states in Fig. 3 of the main text. For the tilted 25-cluster and the 20-site cluster (which is also tilted), only one of the patterns is compatible with periodic boundary conditions, and it is directly visible in the real-space correlations (see Fig. 2 of the main text). By contrast, the ($5 \times 5$) 25-site cluster is compatible with both patterns. As a consequence, the real space correlations corresponds to a superposition of the two patterns, as can be seen in Fig.2.

## FULL LOW-ENERGY SPECTRUM OF $SU(5)$ ON THE SQUARE LATTICE FOR THE 20-SITE CLUSTER.

To further demonstrate the power of the present approach, we briefly discuss the full low-energy spectrum of $SU(5)$ on the square lattice for the $n = 20$-site cluster. There are 350 different Young tableaux of $n = 20$ boxes and no more than 5 lines. Each of them corresponds to an irrep of $SU(5)$, and we have been able to calculate the lowest ten eigenenergies of the Quantum permutation Hamiltonian in each of them. We have plotted these energies as a function of the quadratic Casimir $C_2$ in Fig. 3. The quadratic Casimir $C_2$ is defined by $C_2 = \frac{1}{2N}\left\{n(N-\frac{n}{N})+\sum_{i=1}^{i=k}\alpha_i^2-\sum_{j=1}^{j=\alpha_1}c_j^2\right\}$, where the $c_j$ $(j=1,2,...\alpha_1)$ are the lengths of the columns [4]. Different colors correspond to different irreps having the same $C_2$. We have also introduced different symbols for different classes of shapes: i) crosses for shapes with no more than 2 lines, so that these irreps also appear in the study of the $SU(2)$ model on the same cluster; ii) triangles for the shapes with exactly three lines, corresponding to irreps which are also present in the spectrum of the $SU(3)$ model; iii) squares for the four line Young tableaux appearing in the $SU(4)$ model; iv) and finally circles for shapes with exactly five lines. For clarity, we have also reproduced the same plot in the subsequent figures, but with a different shape emphasized in each of them. In the right portion of Fig. 3, one can easily recognize the characteristic $SU(2)$ tower of states. More generally, the full spectrum of an $SU(N)$ model will always contain the energies of $SU(p)$ $(p \leq N)$, so that the full spectrum can be interpreted as the superposition of different spectra. Interestingly, the energies corresponding to $SU(p)$ with the smallest value of $p$ tend to be the lowest energies for the values of the Casimir at which they are present. The only exception is $p=3$, presumably a consequence of the fact that the 20-site cluster is frustrated for $SU(3)$. A detailed analysis of these issues is beyond the scope of this paper and will be the topic of a future publication.

---

\* Electronic address: `pierre.nataf@epfl.ch`

† Electronic address: `frederic.mila@epfl.ch`

[1] A. Young, Proc. London. Math. Soc. **2**, 196 (1931).

[2] R. M. Thrall, Duke Math. J. **8**, 611 (1941).

[3] D. E. Rutherford, *Substitutional Analysis* (Edinburgh University Press, 1948), p. 58.

[4] K. Pilch and A. N. Schellekens, J. Math. Phys, **25** (12) (1994).

[5] P. Corboz, M. Lajkó, A. M. Läuchli, K. Penc, F. Mila, Phys. Rev. X **2**, 041013 (2012).

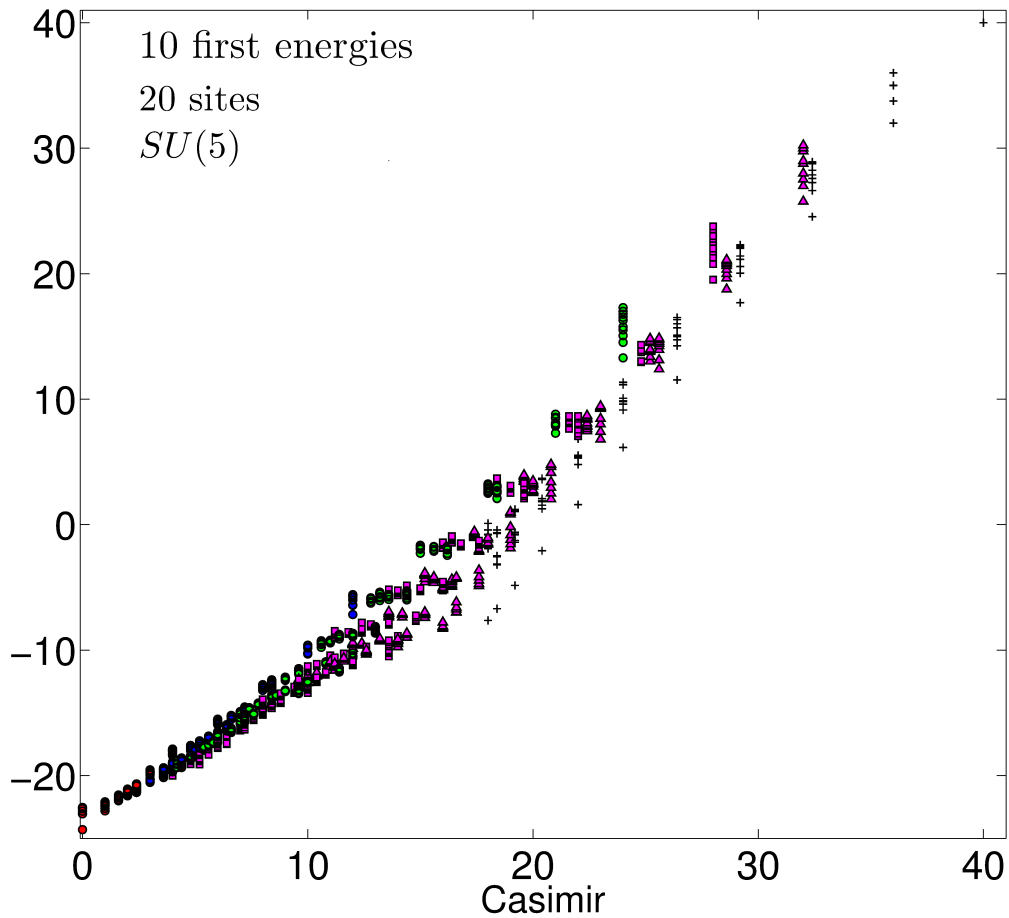

FIG. 3: Full low energy spectra of $SU(5)$ (in units of $J$) for the 20-site cluster. We have plotted the lowest 10 eigenenergies of all the 350 different irreps as a function of the quadratic Casimir $C_2$. Different colors have been used to distinguish different irreps with the same $C_2$. We have also used different symbols for different kinds of shapes $\alpha$: crosses for shapes with at most two rows, triangles for shapes with three rows, squares for shapes with four rows, and circles for shapes with 5 rows. The spectrum can be interpreted as the superposition of different spectra, as emphasized in the next figures.

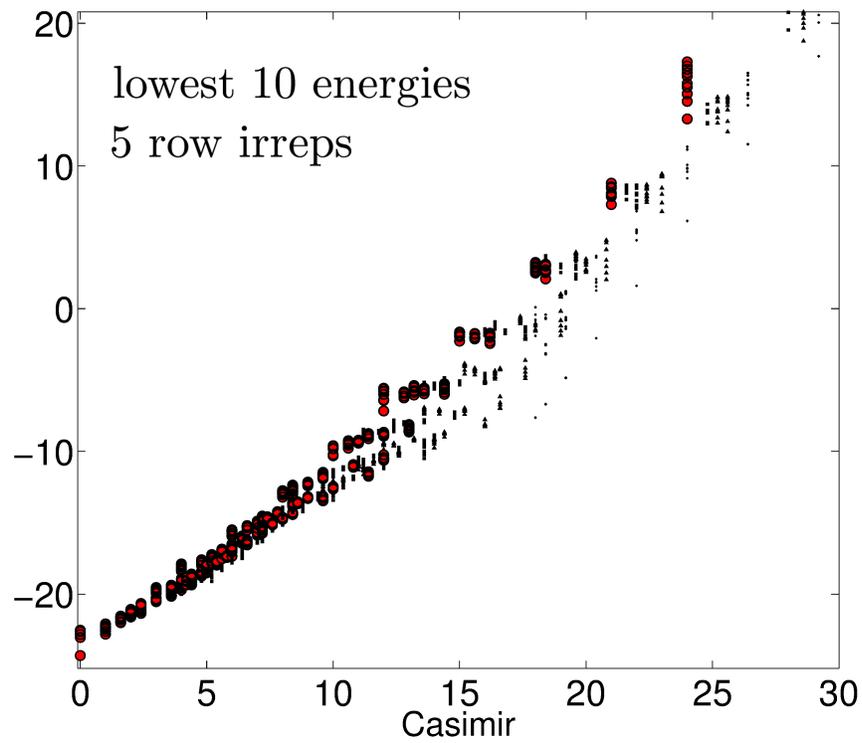

FIG. 4: Same as Fig. 3, but with the energies for the irreps represented by Young tableaux with exactly 5 rows emphasized as bigger red symbols. Note the presence of a tower of states for small $C_2$.

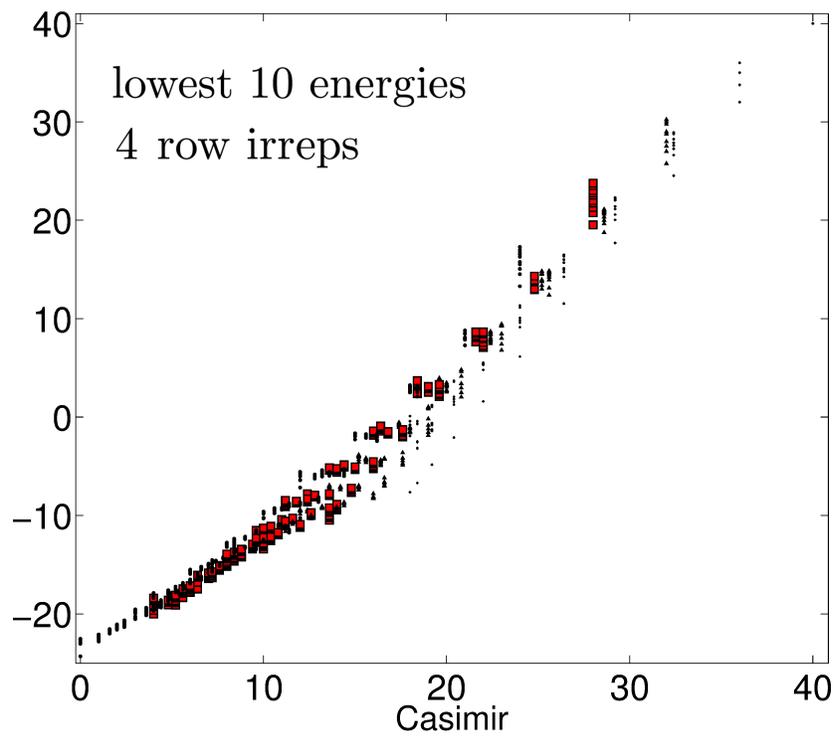

FIG. 5: Same as Fig. 3, but with the energies for the irreps represented by Young tableaux with exactly 4 rows emphasized as bigger red symbols.

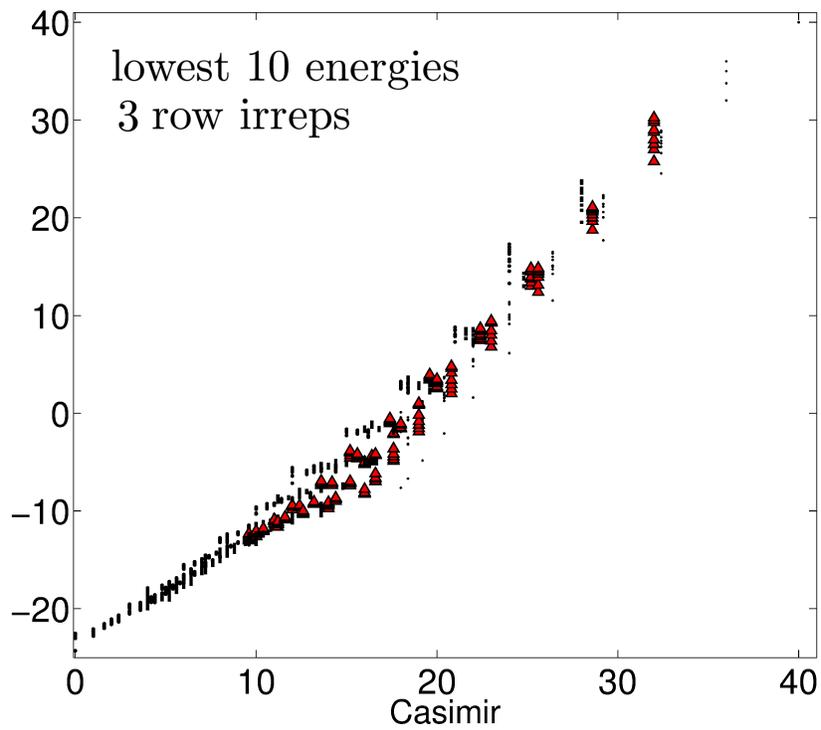

FIG. 6: Same as Fig. 3, but with the energies for the irreps represented by Young tableaux with exactly 3 rows emphasized as bigger red symbols. Note that the 20-site cluster is frustrated for $SU(3)$.

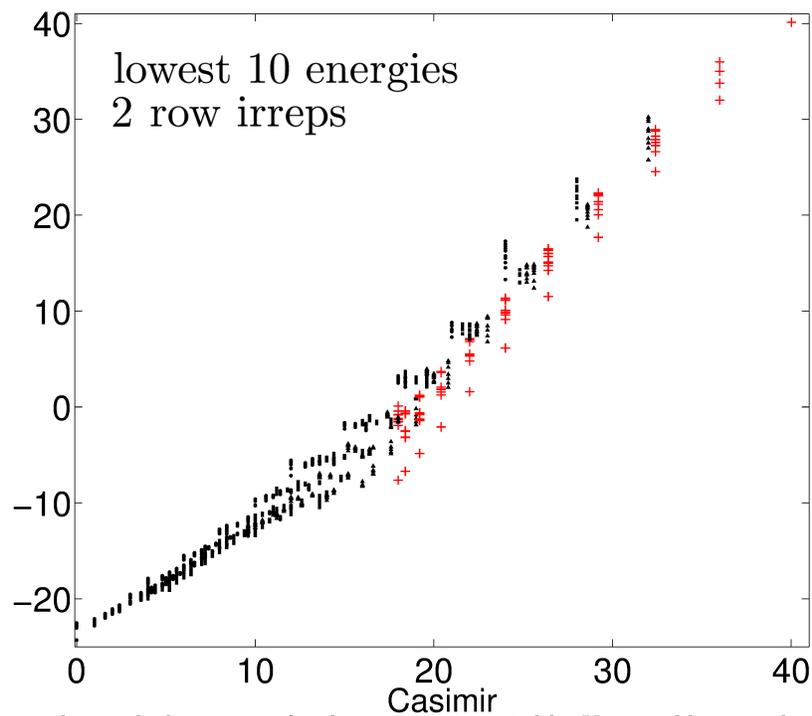

FIG. 7: Same as Fig. 3, but with the energies for the irreps represented by Young tableaux with at most 2 rows emphasized as bigger red symbols. A tower of states for large $C_2$ is clearly visible, a consequence of the fact that the ground state of the Heisenberg $SU(2)$ model on the square lattice has long-range Néel order.